**Single-Shot Phase Diversity Wavefront Sensing in Deep Turbulence via Metasurface Optics**

*Arturo Martin Jimenez[1], Marc Baltes[1], Jackson Cornelius[1], Neset Akozbek[2], and Zachary Coppens[1*]*

[1]CFD Research Corporation, Huntsville, AL, 35806

[2]US Army Space and Missile Defense Command, Huntsville, AL, 35808

[*]Email: zachary.coppens@cfd-research.com

**Abstract:**

Free-space optical communication (FSOC) systems offer high-bandwidth and secure communication with minimal capital costs. Adaptive optics (AO) are typically added to these systems to decrease atmospheric channel losses; however, the performance of traditional AO wavefront sensors degrades in long-range, deep turbulence conditions. Alternative wavefront sensors using phase diversity can successfully reconstruct wavefronts in deep turbulence, but current implementations require bulky setups with high latency. In this work, we employ a nanostructured birefringent metasurface optic that enables low-latency phase diversity wavefront sensing in a compact form factor. We prove the effectiveness of this approach in mid-to-high turbulence (Rytov numbers from 0.2 to 0.6) through simulation and experimental demonstration. In both cases an average 16-fold increase in signal from the corrected beam is obtained. Our approach opens a pathway for compact, robust wavefront sensing that enhances range and accuracy of FSOC systems.

**Main Text:**

**Introduction:**

Free space optical (FSO) communication is a promising emerging technology that employs modulated laser light to transmit data through the atmosphere between two terminals. In contrast to conventional communication technologies, FSO systems offer the combined advantages of high bandwidth, increased security, and readily deployable links with low capital expenditure [1]. Current implementations have shown great promise; however, a major drawback of the technology is that the range and achievable data rates are sensitive to laser beam disturbances induced during atmospheric propagation. These disturbances are especially problematic when propagating long distances through high refractive index gradients in the atmosphere (i.e., deep turbulence).

Adaptive optics (AO) is a technique that can mitigate the effects of atmospheric turbulence and maximize range and accuracy in FSO systems. AO systems typically operate in conjunction with wavefront sensors, which measure the aberrated wavefront of the beam after propagating through the atmosphere. The beam leaving the transmitter is then pre-distorted based on the measurement so that it becomes sharpened during propagation to produce high power/signal at the receiver. The most common wavefront sensor is the Schack-Hartmann (SHWS), which uses a lenslet array to track local slopes of the incoming wavefront (see Figure 1a). While SHWS is appropriate for low turbulent levels, it fails in mid-high turbulent atmospheres due to irradiance fade from scintillation and its inability to find hidden phase from branch points [2,3]. Other wavefront sensors make use of interferometric techniques, such as digital holography, which can accurately measure the wavefront in strong turbulence. Unfortunately, these methods suffer from issues including vibrational sensitivities and doppler shifts from moving terminals, which hinder their practical implementation[4,5].

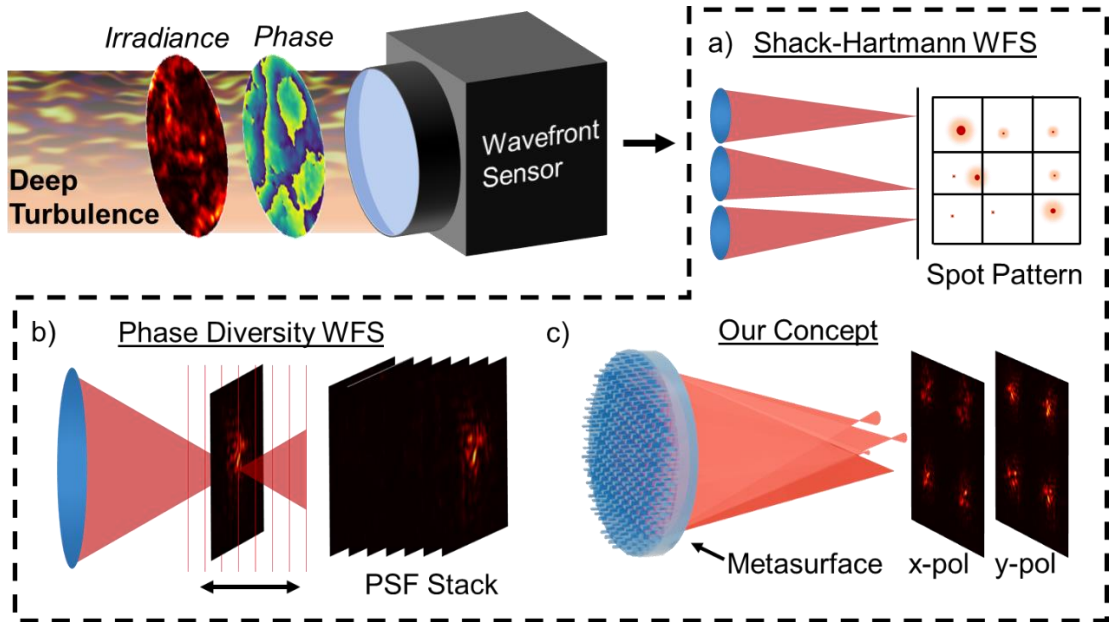

**Figure 1:** Schematic illustration of wavefront sensing techniques. a) Depicts a Shack-Hartmann wavefront sensor using a lens array to capture PSFs whose shift correspond to the local slope of the input wavefront. b) Shows method based on phase diversity where the intensity pattern at various axial positions is captured to solve the inverse problem. c) Shows our method of single-shot phase diversity wavefront sensing using a metasurface to capture multiple PSFs with various focal lengths.

Phase diversity wavefront sensing is an irradiance-based, non-interferometric sensing technique that is tolerant to both branch points and scintillation [6–8] and thus shows effectiveness in deep turbulence. The technique works by capturing intensity profiles at several planes along the system's optical axis, each with a different added phase shift such as defocus (see Figure 1b). The intensity profiles are then fed into an algorithm to solve the inverse problem and recover the amplitude and phase at the entrance aperture. Phase diversity wavefront sensing has been implemented through several methods, such as using beam splitters with multiple offset cameras to capture point spread functions (PSFs) at various defocused planes [6], focusing the input wavefront and scanning the detector through the focus [7], and adding a varying random phase to the input wavefront with a spatial light modulator [8]. Although effective, these implementations often require bulky setups with moving parts that introduce unwanted latency, making them poorly suited for practical applications.

In recent years, there has been an increased research focus on using metasurfaces to reduce the size and complexity of optical systems [9–15]. Metasurfaces are nanostructured materials composed of subwavelength features that allow for precise control of phase, amplitude, dispersion, and polarization in a flat form factor. These unique capabilities have recently been exploited for compact, robust wavefront sensing and phase imaging. A common implementation makes use of polarization-sensitive metasurfaces to create laterally displaced images (shearing imaging) whose interference image can be used to reconstruct the phase and amplitude information [16–18]. This method, however, yields the phase gradient which then needs to be properly integrated to construct the phase. The integration can take several seconds, making it impractical in rapidly changing atmospheres, i.e., high Greenwood frequencies. Other implementations use polarization multiplexing to simultaneously capture two intensity images which are used to calculate the phase with the transport-of-intensity equation (TIE) [19]. This approach, however, can only capture two images which limits the accuracy of reconstruction. Moreover, the TIE approach assumes a uniform intensity distribution and has been shown to not hold for inhomogeneous intensities [20,21], thus precluding its use for reconstructing scintillated wavefronts. While these efforts establish a precedent for wavefront sensing using metasurfaces, a compact, low-latency solution for wavefront sensing in deep turbulence is still lacking.

In this work, we propose and experimentally demonstrate single-shot wavefront sensing in highly turbulent atmospheres using a single-layer metasurface for phase diversity encoding and a convolutional neural network (CNN) for wavefront reconstruction. Our metasurface leverages both spatial and polarization multiplexing to simultaneously generate eight, varied PSFs at the detector. The CNN is trained on these PSFs using an end-to-end approach where the figure of merit (FoM) is taken to be the power in bucket (PiB) obtained at the receiver after propagating the predicted wavefront through the turbulent atmosphere. Yin et al. has shown PiB to be a well-suited metric for evaluating the effects of atmospheric turbulence on the bit error rate (BER) of an FSO system[22]. Our model test results show high accuracy reconstruction predictions in deeply turbulent atmospheres, even in the presence of strong scintillation, branch points, and noise. A metasurface encoder was fabricated and the CNN model was retrained on the experimental captures, again providing high-accuracy reconstruction on unseen wavefronts.

**Results:**

Our wavefront sensor concept is shown in Figure 1c where the metasurface generates eight PSFs at the detector that are separated through spatial and polarization channels. Each PSF is slightly out of focus with respect to one another, providing the same phase diversity in a single capture as would be obtained by translating the sensor along the optical axis. To achieve this functionality, the metasurface is designed to include a supercell architecture with birefringent nanopillars as depicted in Figure 2a. The supercell is composed of four unit-cells, each acting separately to focus light with a different focal length to each quadrant on the image plane. In each unit cell, the nanopillars have varied widths along the x and y axes to provide different phase delays for orthogonal polarization components of the incident light. This birefringent property is exploited to create an additional PSF channel, expanding the four supercell PSFs to eight. The entire focal length stack was constrained between 48.8 mm to 51.6 mm to obtain an f-number of 10 for a clear aperture of 5 mm. This ensured that the PSFs were large enough to be finely resolved by the sensor but not too large so as to bleed over into adjacent quadrants. The defocus range of 2.8 mm was chosen to provide sufficient feature variation among the PSFs for the CNN algorithm.

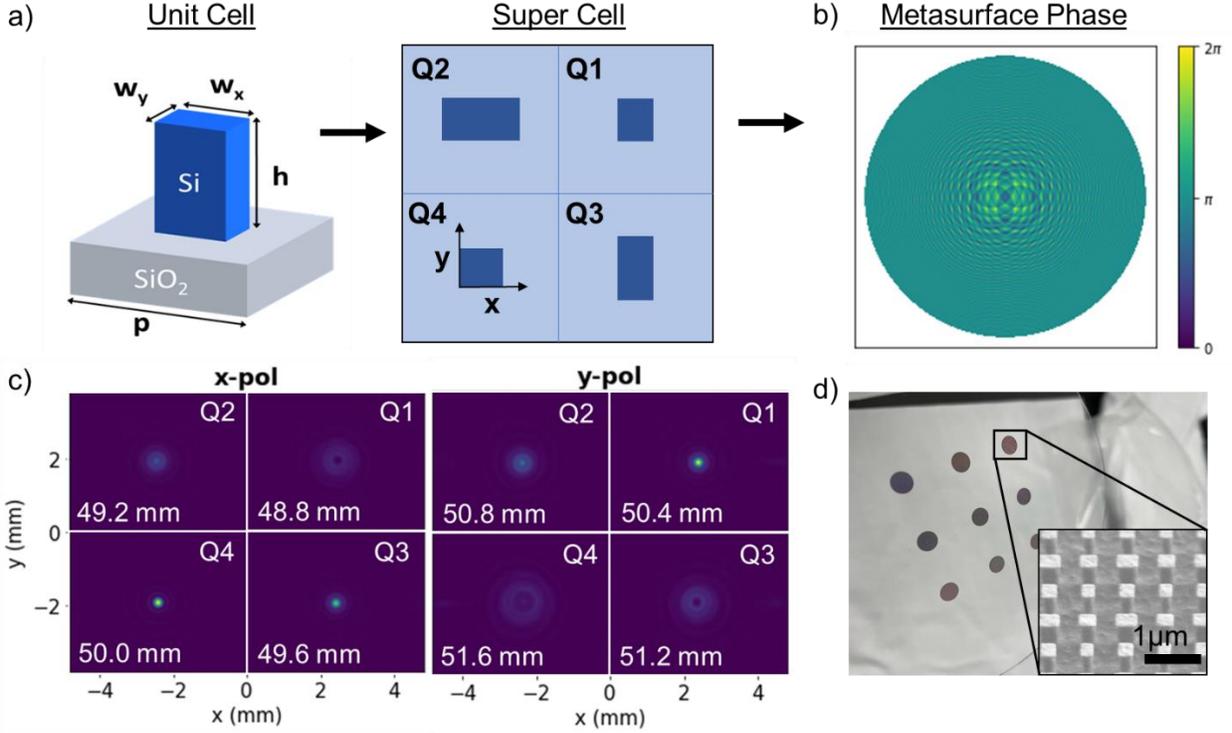

**Figure 2:** Metasurface design and fabrication. a) 3D representation of metasurface unit cell with birefringent silicon pillar on fused silica substrate and schematic of supercell architecture with four quadrants. b) Wrapped metasurface design phase profile for x-pol. c) Simulated design PSFs for orthogonal polarizations (x-pol and y-pol) showing the corresponding focal lengths at each quadrant. d) Image of wafer containing multiple fabricated metasurfaces with the final design. The inset shows an SEM image of the nanopillars.

The design phase, $\varphi$, for each polarization at every position $(x_{ij}, y_{ij})$ in the metasurface was calculated using Equation 1,

$$\varphi(x_{ijk}, y_{ijk}) = \frac{2\pi}{\lambda}\left[f_k - \sqrt{f_k^2 + (x_{ijk} - \Delta x_k)^2 + (y_{ijk} - \Delta y_k)^2}\right] \quad (1)$$

$$f_k = f_0 - (4-k)\Delta f, k = 1,2,3,4 \ (x-\text{pol})$$

$$f_k = f_0 + k\Delta f, k = 1,2,3,4 \ (y-\text{pol})$$

$$\Delta x_k = x + (-1)^k \Delta x$$

$$\Delta y_k = \begin{cases} -\Delta y, k < 3 \\ \Delta y, k \geq 3 \end{cases}$$

where $\lambda$ is the design wavelength, $f_k$ is the focal length at every quadrant $k$ of a supercell at position $(i,j)$ calculated from the defocusing step $\Delta f$, and $\Delta x_k$ and $\Delta y_k$ are the offsets along the x and y axes, respectively. The resulting phase mask for x-polarization is shown in Figure 2b. The unit-cells were modeled using rigorous coupled-wave analysis (RCWA) to obtain the phase and transmission at transverse-electric (TE) and transverse magnetic (TM) polarizations for multiple combinations of nanopillar widths along the x and y axes (see Supplementary Section 1). Nanopillars were selected using a nearest neighbor search algorithm of discretized phase while excluding dimensions that provided <70% transmission efficiency. The simulated captures of the PSFs at the design wavelength of 1550 nm are shown in Figure 2c and confirm the ability of the metasurface to achieve single-shot multiplane phase diversity. It is important to note that more captures along the optical axis generally improves reconstruction accuracy for phase diversity wavefront sensing [23,24] and supercells with even more unit-cells can be used to create additional channels as needed.

After verifying the design parameters, the metasurface was fabricated. First, 1.2 µm of amorphous silicon was deposited on a cleaned fused silica substrate. The metasurface write file was compressed and fed into an electron beam lithography tool to expose the nanopillar features in PMMA photoresist. After development, an $Al_2O_3$ hard mask was deposited followed by a liftoff process. The nanopillars were then etched using a standard silicon Bosch process. An image of the fabricated metasurface is shown in Figure 2d. Additional details of the fabrication are included in the methods section.

Simulated Results

The PSFs generated by the metasurface were fed into a U-Net CNN model for the wavefront reconstruction. The U-Net model was originally used for image segmentation[25] but it has also shown success in image reconstruction tasks[26] and wavefront reconstruction[27]. A typical loss function for training U-Net reconstruction models includes mean squared error between the predicted phase and the ground truth [28] While effective in some reconstruction applications, we found reduced performance due to overfitting even at relativity low turbulence levels (see Supplementary Section 3). To improve performance, we implemented an end-to-end (E2E) approach, modeling the full beam propagation and taking the loss function to be the power-in-bucket (PiB) at the receiver. This approach allowed for optimization of the most important metric, increased signal, and improved the generalizability of the model.

A schematic of the training pipeline is shown in Figure 3a. Briefly, the input wavefronts were generated by propagating a beacon source through thirteen random phase screens with power spectral density (PSD) in accordance to the modified Von Karman model [29] for a given turbulence level (see Supplementary Section 2). We report the turbulence level using the Rytov number, which is the variance of the log-amplitude of the scintillated wavefront. After propagating through the discretized atmosphere, the metasurface phase for each polarization is added to the generated, scintillated wavefront and propagated to the focal plane using the Angular Spectrum Propagation (ASP) method (see Methods). The obtained PSFs from each quadrant were properly cropped and fed to the U-Net model as an eight-channel tensor. In the model, the most important features in the input PSFs are extracted and reduced to a low dimensional feature representation via repeated convolutional and pooling operations. This representation is then decoded using deconvolutional operations and skip connections, which provide the decoder with information from earlier layers and help alleviate the vanishing gradients problem. The U-Net model outputs a predicted phase whose conjugate is propagated with an added focusing phase back through the same random screens used during generation (in reverse order) to obtain a focused beam at the receiver. The PiB of the focused beam is calculated within a bucket of diameter equal to the full width at half maximum (FWHM) of a diffraction-limited Airy disk. The loss is then calculated using the normalized PiB (NPiB), defined as the ratio of the calculated PiB to the PiB obtained from the ideal Airy disk.

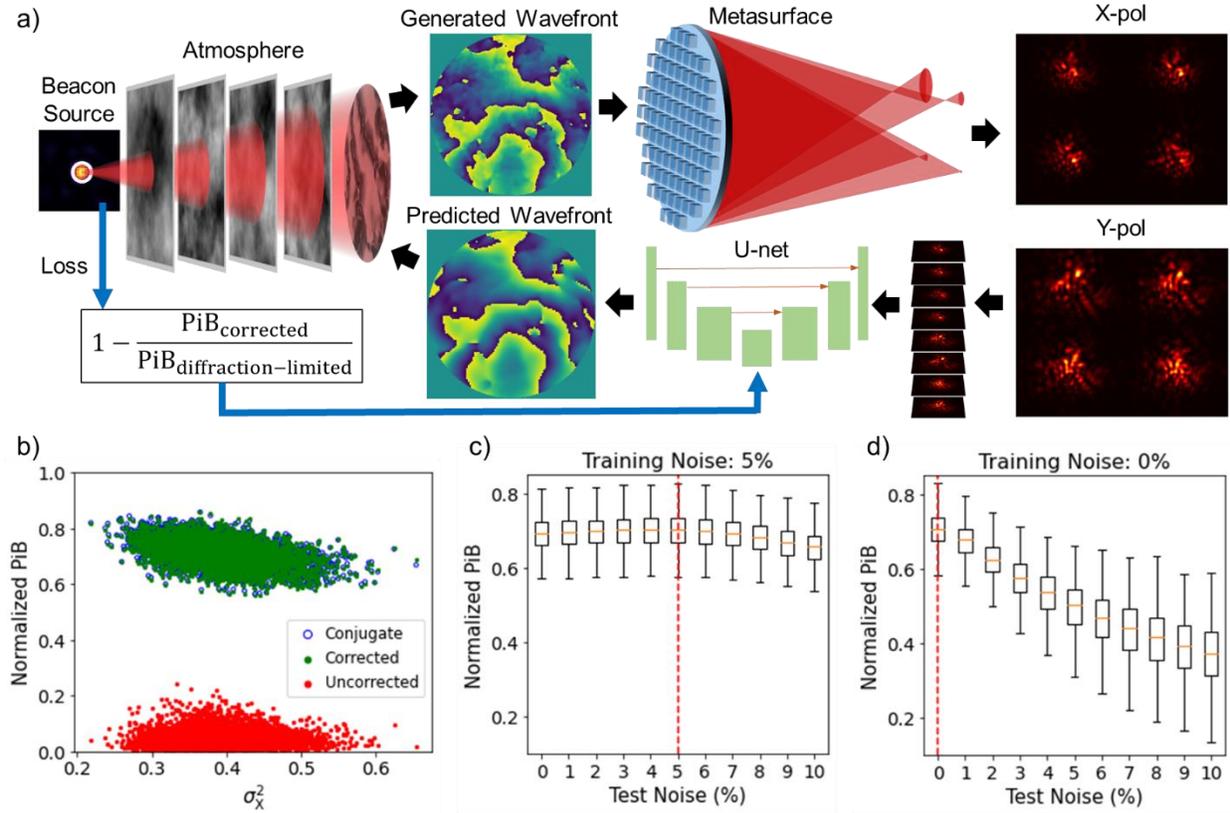

**Figure 3:** Model training pipeline and simulation results. a) Shows the full pipeline used to train the U-Net model, beginning by generating a scintillated wavefront and focusing it with the metasurface, then the captured PSFs are fed to the model to output a predicted wavefront which is backpropagated through the random phase screens simulating the atmospheric turbulence to obtain a PiB. Finally, the loss function is calculated using the PiB and the model weights are updated. b) NPiBs calculated using the conjugate of the generated phase (blue circles), the predicted phase (green dots), and a flat uncorrected wavefront (red dots) plotted vs. the calculated Rytov number for each sample wavefront. NPiBs obtained from inference runs with PSFs containing various noise levels from 0% to 10% are shown c) when using a model trained on PSFs with 5% noise and d) when using a model trained on noiseless PSFs. The dashed red lines indicate the level of noise present in the training data. The orange lines on each boxplot mark the median values.

For the simulated study, 100,000 wavefronts were generated consisting of Rytov numbers between 0.25 and 0.6. For reference, SHWS estimation accuracy significantly degrades beyond Rytov of 0.2, providing almost no accuracy for Rytov of 0.6 when using subaperture diameters near the turbulence coherence diameter[2]. The data was split into training, validation, and testing sets using a 90/5/5 split. This gave the model 90,000 samples for training, 5,000 for validation, and 5,000 for testing. After training for 500 epochs, the model showed good convergence and generalization. Figure 3b shows the calculated NPiBs from post-training inference results when the model is presented with PSFs from the test set. The results show an average 16-fold increase in signal when comparing an uncompensated (uncorrected) beam to the compensated (corrected) beam. Additionally, we show that the NPiB for the conjugate of the incoming wavefront (i.e., the ideal compensated wavefront) nearly matches the NPiB provided by our model, indicating our wavefront sensor is achieving the best possible performance. We note the NPiB does not reach the Airy disk value due to lack of time reversal symmetry, since the back-propagated wavefront has a flat amplitude profile instead of the scintillated amplitude that results from propagating through the atmosphere. This also causes a slight decrease in the corrected and conjugate PiBs for increasing Rytov numbers (see Supplementary Section 2). Generated and predicted wavefronts for a representative high-

Rytov sample are included as insets in Figure 3a. Good qualitative agreement can be observed in the wavefronts, despite the model not being explicitly trained to accurately predict the wavefront. While the prediction does smooth over some of the high phase gradients in the ground truth, this does not diminish the ultimate metric of increased signal. In fact, for some samples, the calculated corrected NPiB is higher than the conjugate NPiB, thus supporting the use of PiB as the proper figure of merit in end-to-end training.

Machine learning (ML) wavefront recovery algorithms are generally more noise tolerant than analytical approaches because ML models can be trained using noisy wavefront data [6,30,31]. To demonstrate this, we added 5% Gaussian white noise to the PSF images and retrained the model. Figure 3c shows boxplots for the NPiB distribution obtained after inference with unseen PSFs containing varied levels of noise between 0% and 10%. The results demonstrate PiB enhancement across a broad noise spectrum, validating the effectiveness of using ML for wavefront sensing algorithms. Interestingly, if a model trained purely on noiseless PSFs is fed noisy PSFs during inference, the NPiB is greatly diminished with increasing noise level (see Figure 3d). This underlines the importance of training on relevant and ideally experimental data when moving ML models from simulation to the real world.

Experimental Results

To experimentally demonstrate the proposed method, the fabricated metasurface was placed in the setup shown in Figure 4a. A fiber-coupled 1550 nm laser was collimated and sent to a spatial light modulator (SLM) to shape the phase of an incident beam. The SLM phases correspond to scintillated wavefronts generated with a nominal Rytov number of 0.35. After reflecting off the SLM, the beam propagates approximately 46 cm to the metasurface. Our simulations showed that after propagating this distance, the beam is scintillated with an average Rytov number of 0.43 (see Supplementary Section 4). After the SLM, a quarter waveplate was used to convert the linearly-polarized light into circularly-polarized light and a linear polarizer was used to discriminate between orthogonal polarization components. A FLIR A6750 S/MWIR camera with a cooled InSb sensor was used to capture the PSFs. PSFs generated with a flat wavefront are displayed in Supplementary Figure S7b and show that the metasurface and the setup are working as intended.

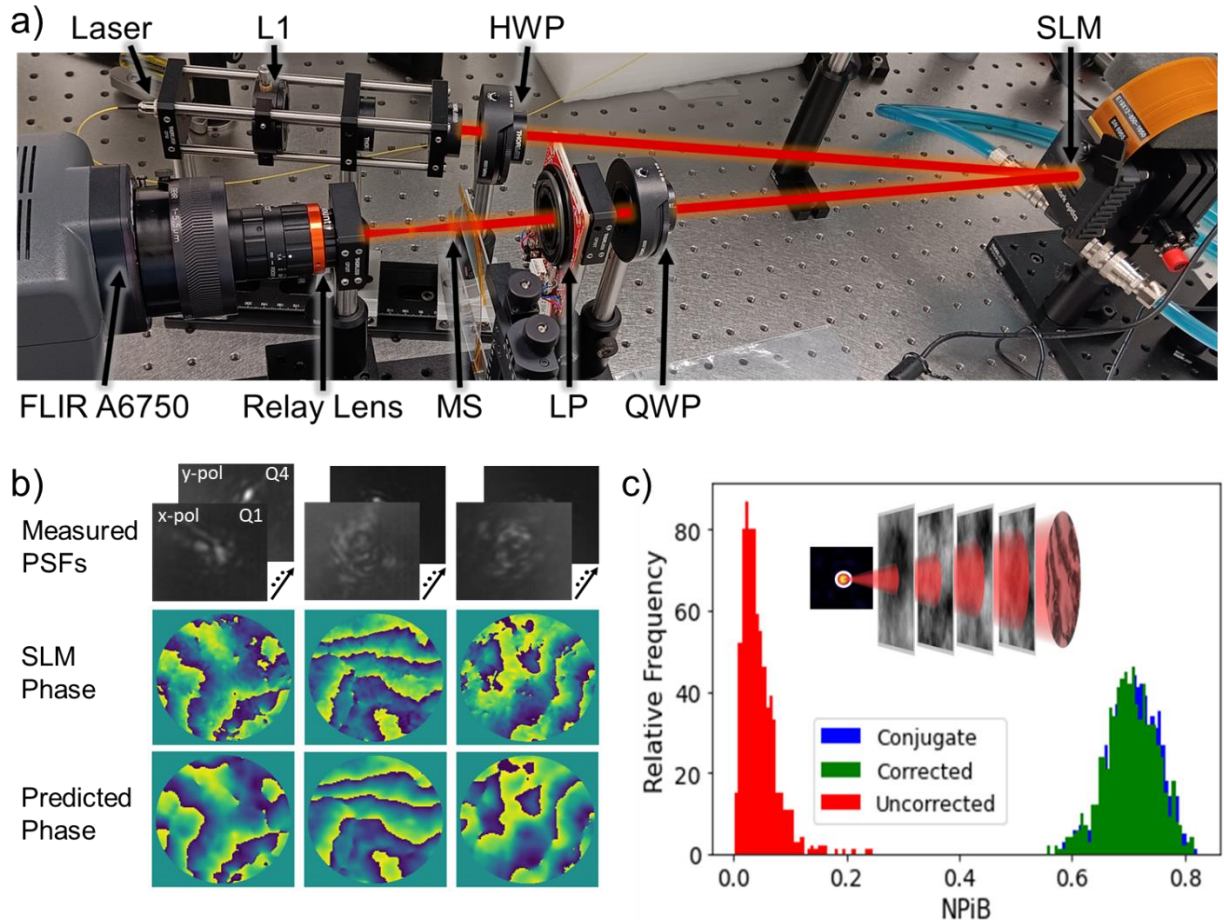

**Figure 4:** Experimental demonstration. a) Image of the experimental setup used, with a 1550 nm pigtailed laser diode, an SLM for phase modulation, a form-birefringent metasurface, an S/MWIR camera and a linear polarizer to discriminate between orthogonal polarizations. b) Shows samples of captured PSFs and their corresponding SLM and predicted phases. A piston phase is added to some of the predicted phases followed by wrapping to [-π,π] for better qualitative comparison. c) Histogram plot of the normalized PiBs obtained for the conjugate (SLM phase), corrected (predicted phase) and uncorrected (flat phase) cases of each sample in a test set after performing transfer learning with the U-Net model using the experimental data.

For the experiment, 15,000 wavefronts were generated with Rytov numbers ranging from 0.30 to 0.62 and the corresponding PSFs were captured. Transfer learning of the U-net recovery model was then performed whereby weights from the previous synthetic training were used to initialize the model for training with experimentally captured data. The training, validation, and testing sets for this experiment were created using the same 90/5/5 split as in the simulated case. The training set consisted of 13,500 samples while the validation and testing sets each contained 750 samples. The model was trained for 500 epochs using the PiB as the FoM as was the case for the simulated data. The PiB was calculated in the same way, by taking the predicted phases and numerically propagating through the random phase screens corresponding to the phase on the SLM. Some of the captured PSFs for test set wavefronts are shown in Figure 4b, along with the generated and predicted phases. The random phase imparted by the SLM produces highly varying features in the captured PSFs. The U-Net model after transfer learning is able to recognize these features and map them to a wavefront phase that yields good qualitative agreement with the generated phase and an increased PiB. In fact, the histogram in Figure 4c shows the average corrected NPiB is 16 times higher than the average uncorrected NPiB. This significant gain is consistent with our simulated results. The corrected

NPiB distribution is also mostly superposed over the conjugate NPiB distribution, showing that the model achieves near-optimal performance. Evaluation for the model was done using an NVIDIA RTX A6000 GPU. The average inference time for a single sample was 2.49 ms, allowing for a frame rate of 401.6 Hz. We note, however, that the model was not optimized for speed and latency could be reduced.

**Discussion:**

While our current implementation shows great promise, there are a number of modifications that can be explored to further improve performance. For instance, in our metasurface design, we chose a clear aperture of 5 mm to expedite the fabrication process, but our design can be easily scaled to larger apertures. Additionally, an f-number of 10 was chosen to prevent the PSFs from bleeding over into adjacent quadrants, but this only holds for Rytov numbers below a certain value. For strong enough turbulence, the PSFs will break down significantly and could cause some of the energy to bleed over into other quadrants. We can compensate for this by changing the focal length values and focus range, which can be optimized when adding the metasurface parameters to the existing end-to-end pipeline. Doing this would allow for co-optimization of the metasurface and the reconstruction algorithm for improved performance, and it could open opportunities for finding phase diversity patterns outside of defocus that can help the model predict wavefronts in increasingly turbulent atmospheres.

There are also areas where the experimental demonstration can be improved. We currently use a linear polarizer to discriminate between the orthogonal polarizations, which requires rotating the polarizer between captures. To make the system operate in a single shot, one can readily implement a polarization-sensitive camera that employs a wire grid polarizer on each pixel of the focal plane array to extract polarization information. This would enable the simultaneous capture of the PSFs for orthogonal polarization without the need for any mechanical movement. To further reduce latency, a dedicated chip can be used to run the U-Net model, allowing for shorter inference times to overcome any Greenwood frequency limitation in deep turbulence.

In summary, we have demonstrated a phase diversity wavefront sensor using a single form-birefringent metasurface in conjunction with a U-Net model which can significantly increase the sensing capabilities of AO systems in the presence of strong atmospheric turbulence. Our implementation is shown to be robust against noise and its simulated and experimentally demonstrated performance surpasses that of traditional wavefront sensing methods in high-Rytov regimes. Our approach opens opportunities for compact, robust wavefront sensing in deep turbulence that enables extended range and accuracy of FSO systems as well as expanded capabilities for other laser-based systems including remote chemical sensing and light detection and ranging (LIDAR).

**Methods:**

Metasurface fabrication

For the fabrication of the metasurface, the material system chosen was silicon on fused silica, as it provides good transmission for the design wavelength and can be etched using Bosh process. An OASIS file was created containing the widths of each resonator at their corresponding positions on the metasurface. This file was sent to the nanofabrication lab at Pennsylvania State University for fabrication. For the final design, a unit cell period of 0.7 µm was used, with a Si pillar height of 1.2 µm and pillar widths ranging from 0.1 µm to 0.5 µm. PECVD was used to create a 1.2-µm-thick layer of silicon on a 500 µm fused silica wafer. Then, electron-beam lithography was used to pattern the silicon layer followed by a deep reactive-ion etching to create the pillar structures.

Numerical Wave Propagation Model

All optical wave propagations were done in Python using the Angular Spectrum[32] method with FFT. Using this method, the propagated field is related to the initial field as shown in Equation 2.

$$U(x, y, z) = \mathcal{F}^{-1}\{M\mathcal{F}\{U(x, y, 0)\}\} \quad (2)$$

$$M = \begin{cases} e^{i2\pi z\sqrt{\left(\frac{n}{\lambda}\right)^2 - (f_x^2 + f_y^2)}}, & \text{for } f_x^2 + f_y^2 \leq \left(\frac{n}{\lambda}\right)^2 \\ 0, & \text{otherwise} \end{cases}$$

This model was used to simulate the focusing of the metasurface and propagate the generated wavefront to the detector plane to obtain the PSFs. Additionally, it was used in the wavefront generation to propagate the point source field through the discretized turbulent atmosphere using split-step propagation. Split-step propagation assumes the effects of propagating the field through a volume with nonuniform refractive index can be captured by doing multiple discrete propagations and adding at each step the accumulated effect of the distance propagated. This split-step propagation was also used during training of the U-Net model to propagate the corrected wavefront back through the discretized atmosphere to the receiver.

Training Details

Each experiment was trained using a U-Net model. The number of filters and kernel sizes in each layer was kept the same those used by Ronneberger et al.[25]. The PSFs were resized to 256x256 and stacked along the channel dimension to create an 8-channel input. Transpose convolutional layers were used for the deconvolutional sections in the network. Once the phase is predicted using the model, it was then masked and shifted to be centered with the aperture.

The models were trained for 500 epochs using the AdamW optimizer with a learning rate of 0.0001 and a weight decay of 0.01. A learning rate scheduler was also incorporated into the optimizer which reduced the learning rate by a factor of 0.5 every 200 epochs. Additionally, the models were trained on an NVIDIA RTX A6000 GPU with a batch size of 64.

**Data Availability:**

Generated datasets used for training and captured simulated and experimental PSFs are available from the corresponding author upon request.

**Code Availability:**

The code for data generation is available from the corresponding author upon reasonable request.

**Acknowledgements:**

This work was supported by the U.S. Army Space and Missile Defense Command.

# Supplementary Information

## 1. Metasurface Design and Fabrication

The metasurface design process is divided into three tasks: 1) determining the phase profiles required to achieve the desired point spread functions (PSFs) at each polarization, 2) designing the unit cell library that can provide the required phase coverage, and 3) discretizing phase profiles and generating Open Artwork System Interchange Standard (OASIS) file with appropriate unit cell dimensions for fabrication. These tasks are not necessarily performed sequentially and often require multiple iterations between them to achieve a final design. In this work, however, we leveraged previous experience with metasurface design and fabrication to accelerate the design process by choosing appropriate unit cell dimensions for Task 1 that allow for flexibility during Task 2 and do not exceed fabrication limitations. As such, we will discuss these tasks in the aforementioned order.

### 1.1: Phase Profile Designs

As mentioned in the main text, the metasurface was designed using a super-cell architecture with birefringent nanopillars. The super-cell architecture allows for spatial separation of the PSFs on a single focal plane, while the birefringent nanopillars enable the polarization multiplexing. In this way, we are able to create four PSFs per polarization channel for a total of eight PSFs. Each super-cell is composed of four nanopillars, each of which is designed to focus the light to one quadrant on the sensor plane with two different focal lengths corresponding to each orthogonal polarization. For a schematic of the super-cell architecture see Figure 2a in the main text. Here, we will discuss all design parameters used for the design of the metasurface phase profiles for each polarization channel, henceforth denominated x-pol and y-pol.

The phase profiles for the x-pol and y-pol polarizations were calculated using Equation 1 from the main text (shown below for convenience). The $(x_{ijk}, y_{ijk})$ coordinates were created from a discretized space of size $D$ with a period $p$ of 0.7 μm corresponding to the unit cell period used in Task 2. The super-cell period is twice the unit-cell period, so the subscripts $i$ and $j$ span $\left[0,1,2,\dots,int\left(\frac{D}{2p}\right)\right]$, while the subscript $k$ spans [1,2,3,4]. This means that the phase is calculated by iterating through every unit-cell at positions $(i,j)$ and every unit-cell at quadrants $k$ within each supercell. The focal length $f_k$ is different for every quadrant $k$ within the super-cell and for each polarization. It is calculated from the nominal focal length $f_0$ of 50 mm and a defocus term $k\Delta f$, where $\Delta f = 0.4$ mm. This term is applied so that light focuses before or at $f_0$ for the x-pol and after $f_0$ mm for the y-pol, as can be seen from the values on Table S1. Finally, horizontal $\Delta x_k$ and vertical $\Delta y_k$ terms are added to the coordinates to shift the position of the PSFs to their corresponding quadrants. The amplitudes for $\Delta x_k$ and $\Delta y_k$ were determined to be 240 μm and 192 μm, respectively. Using a relay lens with 10X magnification, these offsets position each PSF in the middle of each quadrant on a 640x512 sensor with pixel size of 15 μm.

$$\varphi(x_{ijk}, y_{ijk}) = \frac{2\pi}{\lambda}\left[f_k - \sqrt{f_k^2 + (x_{ijk} - \Delta x_k)^2 + (y_{ijk} - \Delta y_k)^2}\right] \quad (1)$$

$$f_k = f_0 - (4-k)\Delta f, k = 1,2,3,4 \ (\text{x}-\text{pol})$$

$$f_k = f_0 + k\Delta f, k = 1,2,3,4 \ (\text{y}-\text{pol})$$

$$\Delta x_k = x + (-1)^k \Delta x$$

$$\Delta y_k = \begin{cases} -\Delta y, k < 3 \\ \Delta y, k \geq 3 \end{cases}$$

**Table S1:** Focal lengths corresponding to each quadrant on sensor plane for each polarization

| x-pol | | | | y-pol | | | |
|---|---|---|---|---|---|---|---|
| QI | QII | QIII | QIV | QI | QII | QIII | QIV |
| 48.8 mm | 49.2 mm | 49.6 mm | 50.0 mm | 50.4 mm | 50.8 mm | 51.2 mm | 51.6 mm |

1.2: Unit Cell Library Design

The unit cells were modeled using rigorous coupled-wave analysis (RCWA) to obtain the phase and transmission at transverse-electric (TE) and transverse magnetic (TM) polarizations for multiple combinations of resonator widths along the x and y axes. The unit cell was modeled with a periodicity of 0.7 µm, as was used in Task 1, and 1.2-µm-tall Silicon (Si) nanopillars on a fused silica ($SiO_2$) substrate. The phases and transmission for TE and TM polarizations are shown in Figure S1. It can be seen that the $0^{th}$ order transmission for both TE and TM polarizations are indistinguishable from the total transmission, which means that higher-order diffractive modes are successfully suppressed. This is mainly achieved by properly choosing the period to be smaller than the effective wavelength in the substrate. In our case, a period of 0.7 µm was sufficient to obtain the phase coverage needed while minimizing higher-order diffractive mode.

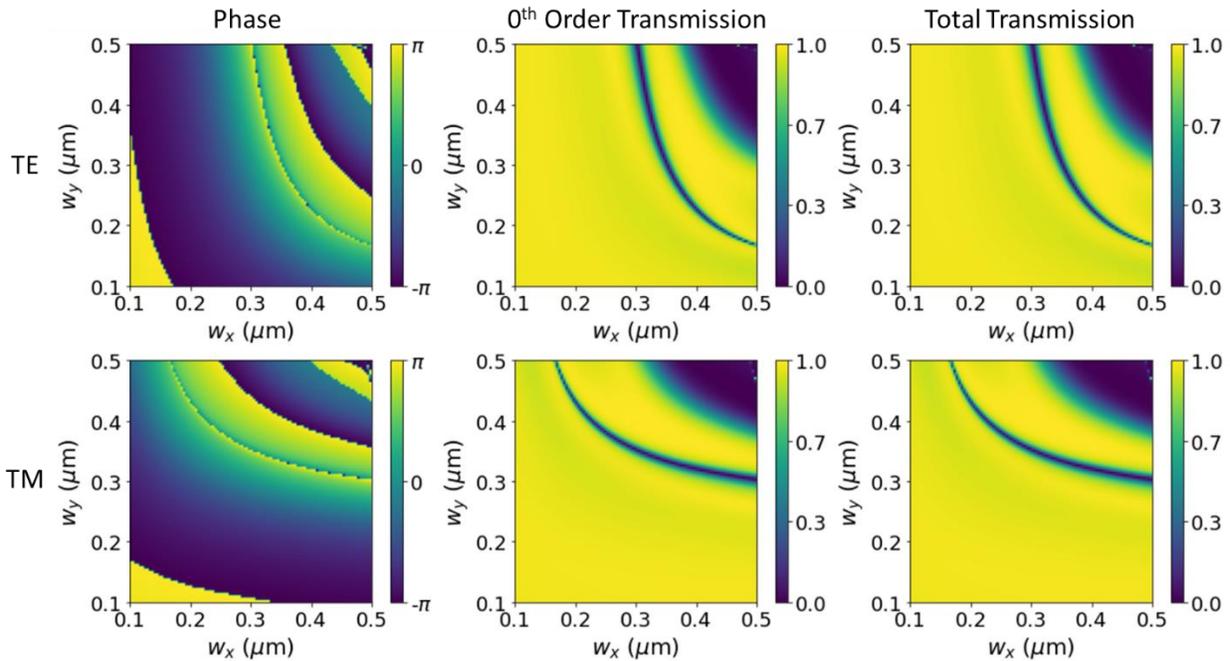

**Figure S1:** Results of RCWA for unit-cell library containing nanopillars with widths ranging from 0.1 to 0.5 microns. Plots on the first and second row correspond to results for TE- and TM-polarized light, respectively. The first column shows the phase delay, between -π and π, at each ($w_x$, $w_y$) combination. The second column shows the transmission efficiency for the $0^{th}$ order diffractive mode, while the third column shows the total transmission efficiency.

There are several wrappings in the TE and TM phase maps, which are necessary to obtain a wider phase coverage. In isotropic designs, a single wrap (2π coverage) is enough to match any design phase profile;

however, due to the anisotropy of our design, our library must satisfy every phase combination from x-pol and y-pol phase profiles at every position on the metasurface. This means that a single phase wrapping for each polarization may not yield sufficient phase coverage. A wider range of resonator widths must therefore be used, which in turn also introduces resonances that significantly reduce the transmission. These resonances must be averted when discretizing the library and choosing the pillar dimensions for fabrication.

1.3: Phase Discretization and OASIS File Generation

The resulting phase coverage is shown in Figure S2a, where the blue dots represent the (TE, TM) phase combinations possible with transmission efficiency above 70% and the red circles represent the combinations selected to discretize the design phase. These red circles are selected by first defining a 16x16 bin grid in the phase coverage map and then selecting the blue dot closest to the center of each bin. Figure S2b shows the discretized ($w_x$, $w_y$) width combinations corresponding to each red circle in the discretized phase coverage plot. With this discretized library, a nearest neighbor search is used to match the (x-pol, y-pol) design phase combination at each position of the metasurface to the closest ($\Phi_{TE}$, $\Phi_{TM}$) combination represented by the red circles. In this way, discretized phase maps and their corresponding width maps are created. Figure S2c shows the obtained discretized phases for the x-pol and y-pol profiles, which are in good qualitative agreement with the design phase profiles (see main text for x-pol profile) and have a discretization root-mean-squared error RMSE of 0.7 radians. The discretized width maps shown in Figure S2d display the discretized width maps which were used to generate the OASIS file for fabrication.

We note that there exist gaps in the phase coverage, so the nearest blue dot for some bins within these gaps may lie in adjacent bins. These gaps were minimized by further increasing the pillar width range; however, further increasing this range introduces additional resonances that reduce transmission. The generated unit-cell library provided a good balance and we were able to achieve low discretization errors (peak-to-valley error PTVE = 0.7 rad)

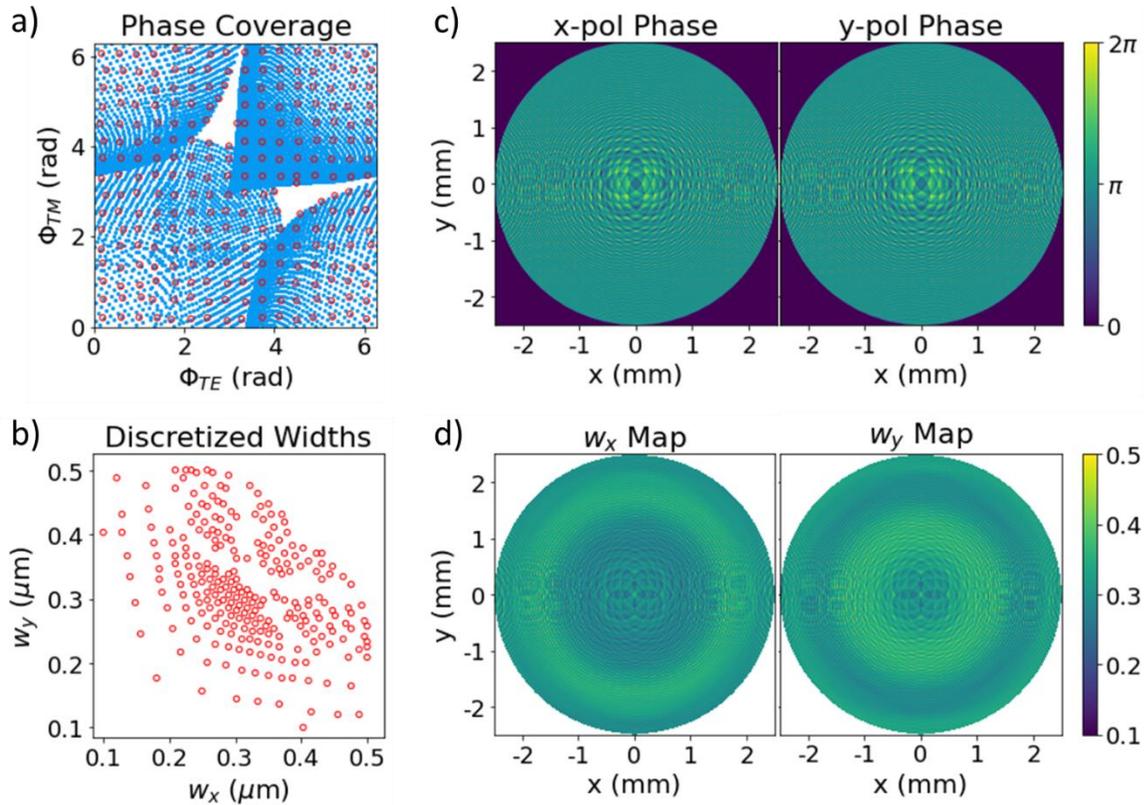

**Figure S2:** Discretization of unit-cell library and design phase profile. a) Shows the phase coverage from the unit-cell library (blue dots) and the selected phase combinations (red circles) to be matched to the design profiles. b) shows the discretized nanopillar widths corresponding to the selected phase combinations in (a). c) The discretized phase profiles at each orthogonal polarization constructed from the selected phase combinations in (a). d) The nanopillar width maps constructed from the widths in (b) corresponding to the discretized profiles in (c).

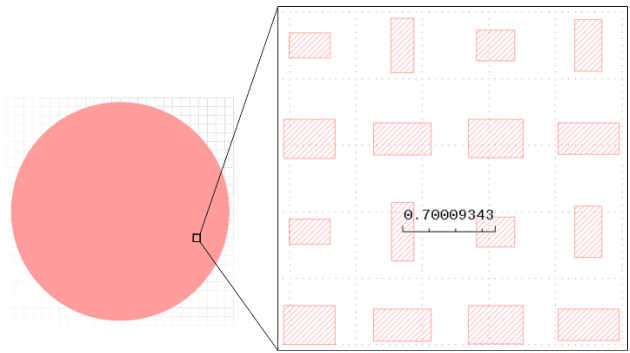

**Figure S3:** The generated OASIS file. The inset shows a closeup of the nanostructures with a period of 0.7 µm.

A rendering of the generated OASIS file using KLayout is shown in Figure S3 with the inset displaying a close-up with a few structures and a measurement to check the period. This measurement was repeated at various positions in the metasurface in both the horizontal and vertical directions. Additionally, the pillar widths at known positions were measured and mapped to their corresponding phase combination from the unit-cell library, which was then compared to the phase combination from the phase profiles at the same position to ensure that the phase difference was within the discretization errors. This was done for pillars at multiple positions on the metasurface before sending the file to the nanofabrication lab at Pennsylvania State University for fabrication.

**2. Wavefront Generation Model**

In this section we will discuss the model used to generate pseudo-random scintillated wavefronts corresponding to specific atmospheric turbulence levels defined by a nominal Rytov number. First, we will describe in detail the process for generating the wavefronts for a given Rytov number, presenting any relevant equations used and explaining their numerical implementations. Then, we will present results from tests used to validate the model accuracy using metrics like the coherence factor, the log-amplitude variance and the long-exposure Modulation Transfer Functions (MTFs). Lastly, we will show results from a brief investigation of the effects of increasing turbulence on the efficacy of the phase correction approach for maximizing power-in-bucket PiB.

2.1 Generating Scintillated Wavefronts

All aberrated wavefronts were generated through numerical propagation (see Methods in main text) of the complex field from a point source through a series of random phase screens that simulate the effects of a turbulent atmosphere. These phase screens where generated using a power spectral density (PSD) profile defined by the Modified Von Karman model for a given coherence diameter (Fried parameter r0). We developed this model based on the equations and sample code found in Jason Schmidt's "Numerical Simulation of Optical Wave Propagation". The process of generating the phase screens is as follows:

1. Define total Rytov number for full propagation distance
2. Discretize turbulence effects and calculate r0 for each phase screen along propagation path
3. Calculate PSD from r0 for each phase screen.
4. Generate 2D array with random numbers from normal distribution
5. Multiply sqrt of PSD to random number array
6. Take inverse Fourier transform

The first step in the wavefront generation process is to define the atmospheric turbulence parameters. The metric used to define turbulence level is the Rytov number ($\sigma_\chi^2$), which is the variance of the log-amplitude of the scintillated field after propagating a certain distance through the atmosphere. Equation S2 shows an analytical relation between the Rytov number and the refractive index structure function $C_n^2$ of the atmosphere through the propagation path (z-axis), where $k$ is the wavenumber, $z_i$ is the position of a random phase screen, $L_z$ is the total propagation distance, and $\Delta z_i$ is the distance between adjacent phase screens. Increasing Rytov number values correspond to higher $C_n^2$ values which mean higher variations in the refractive index, indicating increased turbulence levels. While the value of $C_n^2$ can vary along the propagation, for a horizontal path it can be assumed to be fairly constant. This allows us to calculate $C_n^2$ from a nominal Rytov number using Equation S2. Then using Equation S3, we can calculate the total coherence diameter $r_0$ for the full propagation length. This parameter will be used, along with the nominal Rytov number, to discretize the atmospheric effects into a few propagations through random phase screens.

$$\sigma_\chi^2 = 0.563 k^{7/6} C_n^2 \sum_{i=1}^{N} [z_i(L_z - z_i)]^{5/6} \Delta z_i \qquad (S2)$$

$$r_0 = \left[0.423k^2 C_n^2 \sum_{i=1}^{N} (z_i/L_z)^{5/3} \Delta z_i\right]^{-3/5} \tag{S3}$$

For the second step, we use the relation between $r_0$ and $C_n^2$ shown in Equation S4, to rewrite Equations S2 and S3 as Equations S5 and S6, where now the two parameters of interest ($\sigma_\chi^2$, and $r_0$) are in terms of the coherence diameter $r_{0_i}$ of each individual random phase screen along propagation. Assuming a uniform $\Delta z_i$, one can solve for the values of $r_{0_i}$ that satisfy Equations S5 and S6 to get the turbulence level necessary for each random phase screen along the propagation path.

$$r_{0_i} = [0.423 k^2 C_n^2 \Delta z_i]^{-3/5} \tag{S4}$$

$$\sigma_\chi^2 = 1.33 k^{-5/6} \sum_{i=1}^{N} r_{0_i}^{-5/3} [z_i(L_z - z_i)]^{5/6} \tag{S5}$$

$$r_0 = \left[\sum_{i=1}^{N} r_{0_i}^{-5/3} (z_i/L_z)^{5/3}\right]^{-3/5} \tag{S6}$$

After calculating the appropriate coherence diameter values, the random phase screens can be generated using the Modified Von Karman PSD as in Equation S7.

$$\Phi_{n_i}(f) = 0.023 r_{0_i}^{-5/3} \frac{\exp(-f^2/f_m^2)}{(f^2 + f_0^2)^{11/6}}, \quad 0 \leq f < \infty \tag{S7}$$

here, $f$ is the ordinary spatial frequency, $r_{0_i}$ are the calculated coherence diameters for each screen, $f_m = 5.92/2\pi l_0$ and $f_0 = 1/L_0$, where $l_0$ and $L_0$ are the inner and outer scales whose values were chosen to be 1 cm and 100 m, respectively. A plot of a sample PSD in log-space is shown in Figure S4. The next step is to generate two 2D arrays with random values drawn from a normal distribution for each phase screen. These arrays correspond to the real and imaginary components of a complex number, which is multiplied by the square root of the PSD to create the spectra of the phase screens. Finally, an inverse Fourier transform is performed on these spectra and the real component is kept as the random phase screens to be used during propagation.

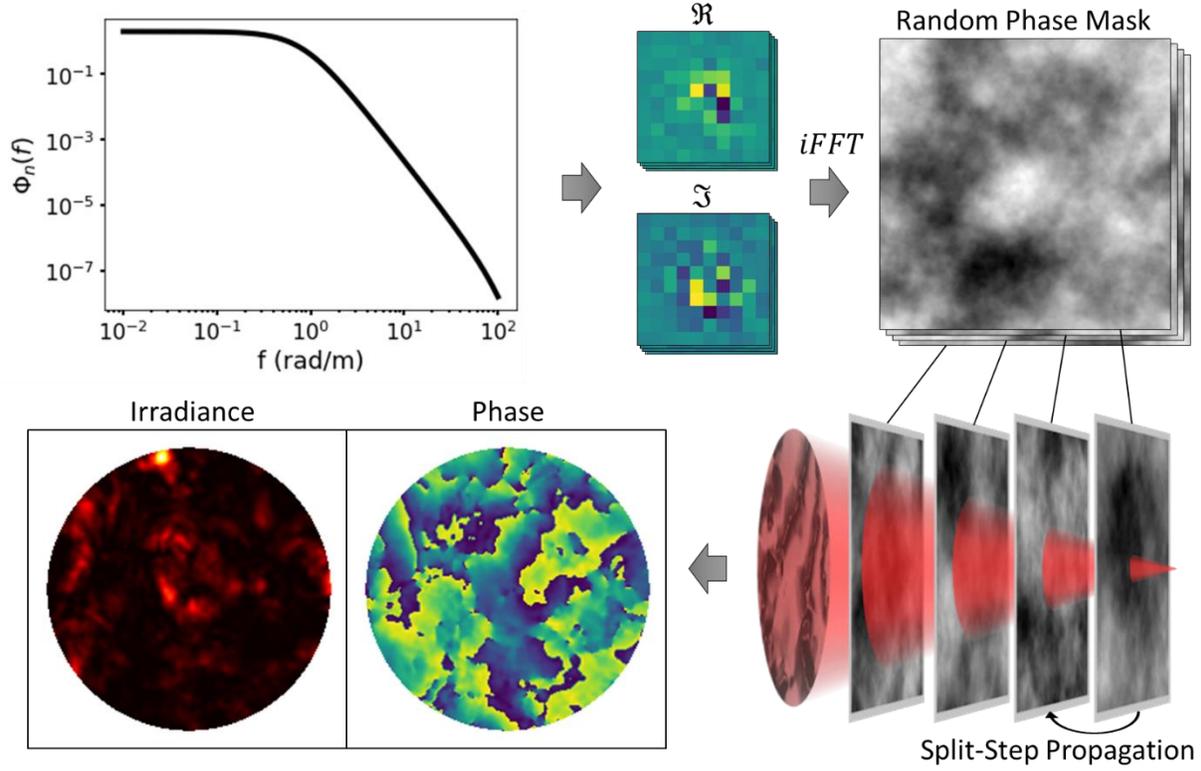

**Figure S4:** Pipeline for generation of scintillated wavefront. The plot at the top left shows a PSD for $r_{0_i} = 0.1$ in a logarithmic scale. This PSD is multiplied to two 2D random arrays to create the real and imaginary components shown to the of the plot. Then, the inverse Fourier transform of these components is taken and taking the real part gives the random phase screen shown at the top right. These phase screens distort the field from the point source through the split-step propagation and the final product is the field shown on the bottom left, with a scintillated irradiance and an aberrated phase.

Once the phase screens are generated, the field from the point source is propagated using split-step propagation from screen to screen until reaching the transceiver telescope. It is important to note that careful considerations must be made for the sampling dimensions and propagation parameters to avoid aliasing while ensuring appropriate resolution to properly capture atmospheric effects. For a more detailed analysis on sampling requirements we refer the reader to Schmidt's book.

After the split-step propagation through the random phase screens, a spherical phase is added to the propagated field to collimate the light. The metasurface phases for each polarization are then added to the collimated wavefront and then each polarization is propagated 50 mm to the focal plane. The captured PSFs, collimated wavefront, and random phase screens are saved for each sample to be later used during training.

2.2 Model Validation

In Figure S5 we show the results of our validation tests. The first parameter measured was the log-amplitude variance, $\sigma_\chi^2$. The box and whisker plots in Figure S5a show the distribution of the calculated $\sigma_\chi^2$ from the generated wavefronts vs. the nominal $\sigma_\chi^2$. The blue dots represent the ensemble average taken over all independent samples generated with a common nominal $\sigma_\chi^2$. Due to the pseudo-random generation of the phase screens, the simulated values can vary about the nominal, but good agreement can be seen between the ensemble average of the independent distributions and the nominal $\sigma_\chi^2$, with a slight deviation becoming more prominent for higher nominal $\sigma_\chi^2$. This is due to the limitation imposed on individual phase screens to not have a $\sigma_{\chi i}^2$ value exceeding 0.1. With 13 screens, of which two (the one at the source, and the one at the final plane) have negligible effect, this limits us to $\sigma_\chi^2 = 1.1$, but we can see that the calculated values start deviating from the nominal values much sooner than that. For our purposes, such high values were not of interest, thus the current accuracy was sufficient.

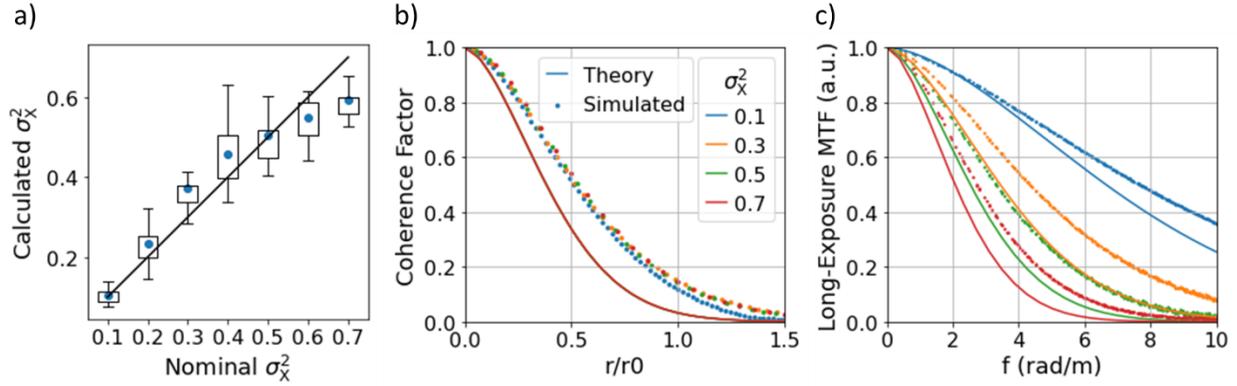

**Figure S5:** Model validation results. a) Shows the calculated vs. nominal $\sigma_\chi^2$ for multiple generated wavefronts. The boxplots represent the distribution of calculated $\sigma_\chi^2$ for generated wavefronts using a single nominal $\sigma_\chi^2$ value while the blue dots are the ensemble average of each distribution. b) Shows the coherence factor averaged over 40 instances of four nominal $\sigma_\chi^2$ values plotted vs. the radial coordinates normalized by the coherence diameter $r_0$. c) Shows the MTFs calculated from the averaged PSFs over 40 instances of four nominal $\sigma_\chi^2$ values (0.1, 0.3, 0.5, 0.7).

The other parameter measured for validation is the modulus of the coherence factor shown in Figure S5b vs. the radial coordinates $r$ normalized to the coherence diameter $r_0$. This was calculated for several independent instances of four nominal $\sigma_\chi^2$ to capture ensemble statistics at each. It can be seen that the calculated coherence factors (scatter dots) are in close agreement with the theoretical values (solid lines). The gap is attributed to the lack of subharmonics in the PSDs used to generate the random phase screens. These subharmonics are responsible for low-frequency effect like tip and tilt whose correction are typically done apart from the higher-frequency corrections, and thus were not of interest for our study.

Finally, we compared the long-exposure modulation transfer function (MTF) obtained from several instances of generated wavefronts for the same nominal Rytov numbers as those used for the coherence factor calculations. The MTFs were calculated by focusing the generated fields to get the point spread functions (PSFs) and then taking the amplitude of the resulting complex array after doing a Fourier transform of these PSFs. Once again, the calculated MTFs shown in Figure S5c are very close to the theoretical values with a small gap which can also be attributed to the lack of subharmonics.

2.3 Increasing PiB in Deep Turbulence

The results shown in Figure 3b of the main text raise an important question which we have briefly investigated in this section. The conjugate power-in-bucket (PiB) is shown to decrease with increasing Rytov number. This PiB is that calculated by backpropagating a wavefront with the negative phase of the generated wavefront, but with a constant intensity profile. In other words, it represents the results that would be obtained in the case of perfectly accurate phase prediction. However, the average PiB obtained with this wavefront is far from diffraction limited, and continues to decrease for higher $\sigma_\chi^2$. As explained in the main text, this is due to a lack of time reversal symmetry. The generated wavefront has a scintillated irradiance profile, while the wavefront that is backpropagated has a flat-top irradiance profile. For increasing $\sigma_\chi^2$, scintillation becomes more severe, and thus it affects the attainable PiB more significantly. To test this reasoning, we performed the backpropagation with phase-only as well as complex-field correction. The latter consisting in propagating the complex conjugate of the generated field. Figure S6 shows the normalized PiBs for both of the aforementioned cases. The boxplots correspond to the distributions of independent generations for seven nominal values of $\sigma_\chi^2$ from 0.1 to 0.7. It can be seen that the maximum PiB possible from the phase-only correction decreases for higher $\sigma_\chi^2$ values, while the complex field correction remains relatively constant with an average value at 95% of diffraction limit with only an increase in standard deviation for higher $\sigma_\chi^2$ values. It is noted that the PiBs obtained from the complex-field correction vary from the diffraction limit, at times being higher. While we have not investigated this in depth, we believe it to be due to two factors: 1) losses from the super-gaussian attenuating mask used to prevent possible aliasing effects, and 2) Numerical errors arising from intensity fluctuations within a small bucket after propagation of highly scintillated wavefronts through randomly generated phase masks.

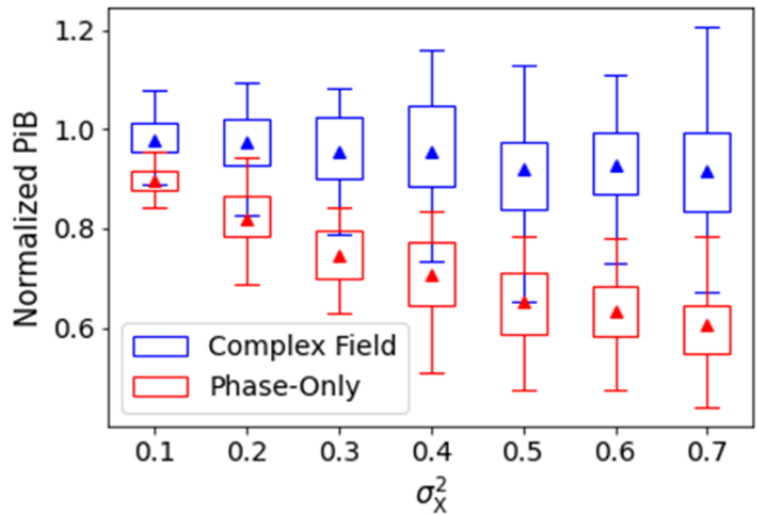

**Figure S6:** Correction method comparison. The boxplots represent the PiB distributions obtained at each nominal $\sigma_\chi^2$ value when propagating a corrected field back to the source. The blue boxplots correspond to fields corrected using the complex conjugate while the red boxplots correspond to fields corrected using only the phase. The triangles show the mean of each distribution.

### 3. Loss Function Selection

We conducted a simulation study to compare two different loss functions: *1)* mean squared error (MSE) between the generated and predicted phase and *2)* PiB ratio of prediction and theoretical maximum using the end-to-end (E2E) method. We chose a nominal Rytov of 0.1 and 15,000 samples to decrease the training time. Our batch size was set to 64 and we split the dataset as follows: 13,500 training, 750 validation, 750 testing. The model was trained for 500 epochs before performing inference on the test set. The results are shown in Table S2. We compared the performance using calculated PiB ratio as this is the ultimate metric for increasing signal in free space optical communication systems. The E2E model provides similar PiB ratios for the test and train data, indicating good generalizability. The MSE model, on the other hand, performs well on the training set but poorly for the test set, indicating overfitting. The results of this study motivated our use of the superior E2E model in the main text.

**Table S2:** Comparison of training and test results for different loss functions

| Model | Split | PiB Ratio |
|---|---|---|
| E2E | Train | 96.3% |
| | Test | 96.0% |
| MSE | Train | 94.6% |
| | Test | 71.5% |

## 4. Experimental Demonstration

The metasurface design and fabrication were validated by our initial measurements in the lab. The experimental setup shown in Figure S7a is the same as that shown in Figure 4a of the main text. A flat phase profile was imparted by the SLM to function as a mirror and the PSFs created by the metasurface were imaged onto the FLIR A6750 camera using a relay system composed of a 6-mm-focal-length relay lens and the bayonet focusing lens from the camera whose focus was adjusted to roughly obtain a 10X magnification. The x-pol was determined first by rotating the linear polarizer in the setup until the observed PSFs looked like the simulated x-pol PSFs, then the y-pol position was 90 degrees from the x-pol. The captured PSFs for each position are shown in Figure S7b along with the simulated PSFs demonstrating good qualitative agreement between the two. This proved that the metasurface was working properly and we could proceed to generate the experimental data by varying the phase on the SLM.

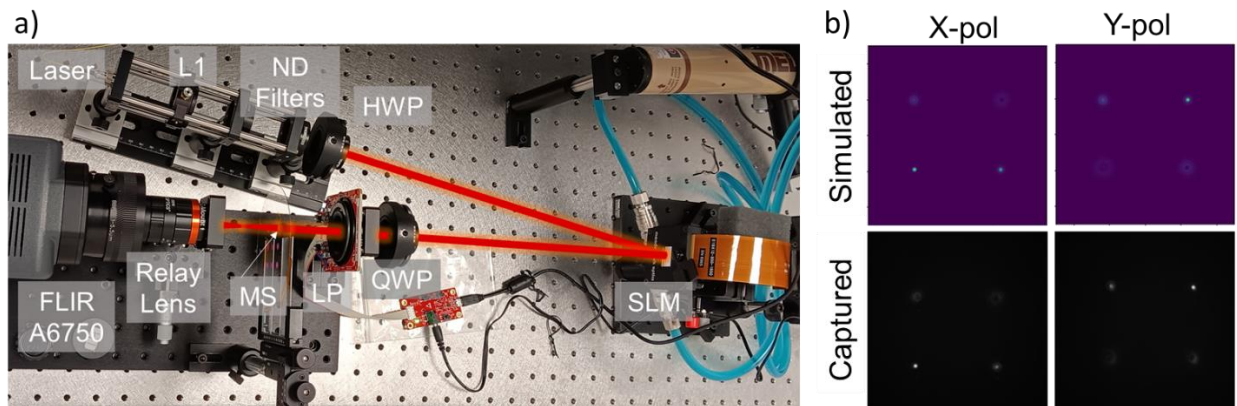

**Figure S7:** Experimental validation of metasurface design. a) Shows the experimental setup used. b) Shows the simulated and experimentally captured PSFs for each polarization.

To generate scintillated wavefronts, a phase was written on the SLM calculated using the model discussed in Supplementary Section 2 for a nominal Rytov number of 0.35. The scintillated beam was simulated by numerically propagating a Gaussian beam, as seen on Figure S8a, after reflecting off of the SLM with the imparted phase. The beam was propagated 46 cm, which is the approximate distance in the lab from the SLM to the metasurface. The Rytov number was then calculated as the variance of the log-amplitude of the propagated beam over a masked aperture simulating the metasurface aperture. This mask was taken to be at a shifted position corresponding to the measured transverse position of the metasurface in the lab with respect to the SLM. This position was determined by using a masked striped pattern on the SLM with phase gradient maximizing the 1st order diffraction mode and varying the center and size of this mask to minimize the signal on the camera. A sample scintillated irradiance is shown in Figure S8b and the distribution of simulated Rytov numbers calculated are shown in Figure S8c using a centered mask and the determined shifted mask. These calculations show that we are testing our wavefront sensing concept in high Rytov regimes, within the range of our simulated results.

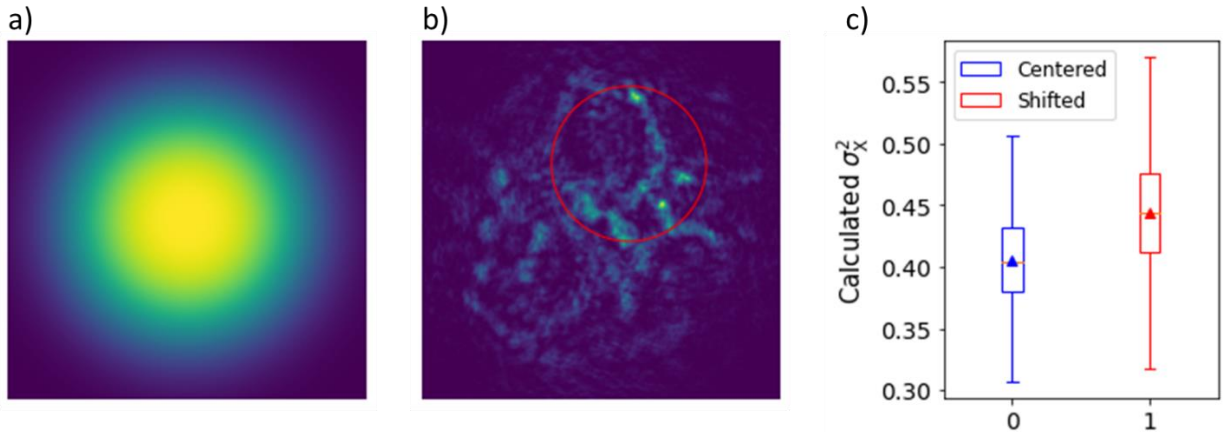

**Figure S8:** Experimental validation of metasurface design. a) Shows the Gaussian intensity profile used to simulate collimated laser beam. b) Shows the resulting scintillated intensity at the metasurface pupil obtained after propagation with the red circle indicating the metasurface aperture mask. c) Shows box plots of the calculated Rytov numbers using a centered aperture (in blue) and the shifted aperture corresponding to the metasurface (in red). The triangles and the orange lines, denote the average and median values of each distribution, respectively.